\documentclass[fleqn,usenatbib]{mnras}

\usepackage{newtxtext,newtxmath}

\usepackage[T1]{fontenc}

\DeclareRobustCommand{\VAN}[3]{#2}
\let\VANthebibliography\thebibliography
\def\thebibliography{\DeclareRobustCommand{\VAN}[3]{##3}\VANthebibliography}

\usepackage{graphicx}	
\usepackage{amsmath}	

\usepackage{slashed}
\usepackage{color}
\usepackage{ulem}

\newcommand{\mBH}{m_{\text{BH}}}

\newcommand{\zvir}{z_{\text{vir}}}
\newcommand{\zobs}{z_{\text{obs}}}

\newcommand{\msun}{M_{\odot}}

\newcommand{\cmg}{\text{cm}^{2}\text{g}^{-1}}

\newcommand{\lrp}[1]{\left({#1}\right)}

\newcommand{\lag}{\mathcal{L}}
\newcommand{\lagdark}{\lag_{\rm{dark}}}

\newcommand{\mchione}{m_{\chi_1}}
\newcommand{\mchitwo}{m_{\chi_2}}

\newcommand{\gchione}{g_{\chi_{1}}}
\newcommand{\gchitwo}{g_{\chi_{2}}}

\newcommand{\chibar}{\overline{\chi}}

\newcommand{\beff}{b_{\rm{eff}}}
\newcommand{\philrd}{\phi_{\rm LRD}}

\newcommand{\mchar}{m_{200}^{\rm{char}}}

\title[Little Red Dot Clustering from uSIDM]{The Clustering of Little Red Dots from Ultra-Strongly Self-Interacting Dark Matter}

\author[Roberts et al.]{
M. Grant Roberts,$^{1,2}$\thanks{E-mail: migrober@ucsc.edu}
Aarna Garg,$^{1}$
Tesla Jeltema,$^{1,2}$
and Stefano Profumo$^{1,2}$
\\
$^{1}$Department of Physics, University of California, Santa Cruz (UCSC),
Santa Cruz, CA 95064, USA\\
$^{2}$Santa Cruz Institute for Particle Physics (SCIPP),
Santa Cruz, CA 95064, USA
}

\date{Accepted XXX. Received YYY; in original form ZZZ}

\pubyear{\the\year{}}

\begin{document}
\label{firstpage}
\pagerange{\pageref{firstpage}--\pageref{lastpage}}
\maketitle

\begin{abstract}

\noindent We predict the effective clustering bias parameter, $b_{\rm{eff}}$, at $z\sim5$ for Little Red Dots (LRDs) seeded by Ultra-Strongly Self-Interacting Dark Matter (uSIDM). From our model, we find that $b_{\rm{eff}}\sim3.6$, thus we infer that LRDs seeded by uSIDM would populate halos of typical masses $\sim 6.3\times10^{10}~M_{\odot}$; this bias factor is consistent with LRDs being a distinct population from high redshift quasars. To the extent that we are aware, this is the first formation-based theoretical prediction of LRD clustering from a model consistent with the LRD mass function. We find that this bias and clustering is insensitive to a wide range of the underlying uSIDM microphysics parameters, including the uSIDM cross-section $\sigma/m$ and uSIDM fraction $f$. This is therefore a robust prediction from the uSIDM model, and will allow for direct probes of the uSIDM paradigm as the origin of LRDs in the next few years. Upcoming \texttt{JWST} observations will constrain the population of LRDs, including directly measuring their clustering.
\end{abstract}




\section{Introduction}\label{sec:introduction}

Little Red Dots (LRDs) are a recently-discovered population of likely dust-enshrouded active galactic nuclei (AGN), most of which have redshifts in the range of $4 < z < 7$; the corresponding super-massive black holes (SMBHs)  are overmassive with respect to local black hole-galaxy scaling relations, and exhibit v-shaped spectral energy distributions, with a sharp Balmer break \citep{Matthee+2024,Kokorev_2024,Greene_2024,Kocevski:2024zzz}. Given the highly compact nature of these objects ($\lesssim$ 100 parsec in size) in addition to the overmassive SMBHs, their existence raises the question of how such massive BHs formed so early in the universe. In particular, LRDs challenge standard models of early galaxy and black hole co–evolution, suggesting unusually rapid or non–standard growth channels at very early times. Moreover, current {\tt JWST} observations increasingly constrain traditional baryonic seeding mechanisms \citep{Volonteri_2010, loeb1994,Inayoshi_SMBH_review,Pacucci+2017,Madau_2014} due to these mechanisms having longer formation time scales, as well as tending to require large periods of super-Eddington accretion to make up for the initial seed formation delay. 

As an alternative approach to seed formation, several papers have considered models of self-interacting dark matter (SIDM) leading to the formation of heavy seeds - either from dissipative effects \citep{DAmico+2017,Latif+2019,Choquette:2019bbi,Xiao_2021,Shen2025,Buckley+2025} or through gravothermal evolution and the resulting core-collapse \citep{Balberg+2002, Pollack:2015iug,Feng2021,Outmezguine:gravothermal,Gad-Nasr+2024,dark_bondi,Roberts_uSIDM,Roberts_LRD}. In particular, ultra-strong self-interactions can rapidly accelerate the gravothermal process in very early halos to form much heavier BH seeds compared to baryonic seeding mechanisms.

In this Letter, we use the uSIDM model \citep{Roberts_uSIDM,Roberts_LRD,Roberts_uSIDM_pheno} to calculate the clustering of the population of LRDs that are seeded by the gravothermal core-collapse of uSIDM halos in the early universe. A possible underlying particle physics model for the uSIDM  has been explicitly worked out in \cite{Roberts_uSIDM_pheno}. In short, the  model consists of two dark matter particles: one is a standard self-interacting DM candidate, while the second one is the ultra-strongly self-interacting particle, which comprises only a small fraction of the dark matter. The mediator of the self interactions is a massive dark photon. The choice of one particle being ultra-strongly self-interacting ensures rapid gravothermal collapse and early seed formation prior to significant baryonic assembly, thereby providing a natural pathway toward overmassive black holes by $z\sim5$.

Recent work has highlighted the need for observations and theoretical predictions of the clustering of LRDs \citep{Arita+24,Lin+25,Schindler+25,Pizzati+25,Carranza+25} which may help distinguish their origin; see \cite{LRD_review_Inayoshi_Ho} for a recent review about LRDs. \cite{Arita+24} performed cross-correlation analysis of 27 low-luminosity broad-line AGNs with $5 < z < 6$ and found characteristic mass scales of $10^{11.46}~h^{-1}\msun$, which are around 1 dex smaller than luminous quasars at the same redshift. \cite{Lin+25} also looked at a sample of low-luminosity AGNs, and they found a characteristic mass scale of $10^{11}\msun-10^{11.2}\msun$, which reinforces that these faint populations are hosted by more moderate size halos than luminous quasars. While both of these studies include significant overlap with LRDs, they do not perform a dedicated clustering or bias measurement for LRDs as a distinct population.

Further, \cite{Pizzati+25} ran mock-clustering simulations to infer potential LRD clustering measurements from \texttt{JWST}, and find that with $\approx10$ fields, they can constrain the characteristic halo mass scale of LRDs to within $0.1-0.3$ dex. Thus, the accuracy needed to determine LRD clustering is within the reach of upcoming \texttt{JWST} surveys. Importantly, their mock results indicate the LRDs may be low-mass halos with modest duty cycles, raising the possibility that LRDs are a distinct population from quasars. On the other hand, \cite{Schindler+25} identified a single LRD at $z\approx7.3$ in an overdensity of eight nearby galaxies and inferred a correlation length of $r_0 = 8 \pm 2 ~h^{-1}~\rm{cMpc}$, which is consistent with ultraviolet luminous high-redshift quasars at $z\approx6$. However, as this determination is based off of a single source, this cannot be a robust conclusion for the entire LRD population. Similarly, \cite{Carranza+25} probed the local environments of LRDs by looking at a photometrically selected sample in $3<z<10$, and found evidence that LRDs tend to reside in less overdense local environments compared to galaxies of similar mass and redshift. 

Together, these studies indicate the need for more LRD observations with \texttt{JWST} as well as for theoretical determinations of LRD clustering bias to determine the nature of these new objects. In this paper we tackle the latter issue and present the first theoretical calculation of LRD clustering. We assume a uSIDM seeding mechanism which can provide a critical test of heavy-seed formation scenarios.

The remainder of this paper is as follows: in the next section we summarize the uSIDM mechanism for early black hole formation; in Section~\ref{sec:beff} we compute the effective halo bias; in Section.~\ref{sec:conclusions} we summarize and conclude.

\section{uSIDM Phenomenology and Black Hole Formation}

The uSIDM model for early SMBH formation (see \cite{Roberts_uSIDM_pheno} for more details) is an extension of the typical SIDM model which also includes an additional, much more strongly self-interacting dark matter species comprising a small fraction, $f$, of the total dark matter budget. The governing Lagrangian is given by,

\begin{eqnarray}
 &&\lagdark = \overline{\chi}_{1}\lrp{i\slashed{\partial} - \mchione}\chi_{1} + \overline{\chi}_{2}\lrp{i\slashed{\partial} - \mchitwo}\chi_{2} \\
 &&~~~~~~~~~~~~~ - \gchione\chibar_{1}A'^{\mu}\gamma_{\mu}\chi_{1} - \gchitwo\chibar_{2}A'^{\mu}\gamma_{\mu}\chi_{2} -\frac{1}{4}F'^{\mu\nu}F'_{\mu\nu} \nonumber\\
 &&~~~~~~~~~~~~~- \frac{1}{2}m_{A'}^{2}A'^{\mu}A'_{\mu}\nonumber ~,
\label{eq:model-lagrangian}
\end{eqnarray}

\noindent where $\chi_1$ and $\chi_2$ are the uSIDM and SIDM particle respectively, and $A^{\prime}$ is the dark photon. In order to have the onset of core-collapse occur shortly after our virialization redshift, $\zvir\sim13.5$, $\chi_1$ couples much more strongly to the dark photon than $\chi_2$, i.e., $\gchione >> \gchitwo$. This guarantees $\chi_1$ to have a significantly enhanced cross-section. From previous uSIDM work \citep{Roberts_uSIDM}, the uSIDM cross-section is on the order of $10^{4}~\cmg$ with a corresponding abundance fraction around $10^{-3.5}$ in order to seed high-redshift black hole formation. 

In \cite{Roberts_LRD} we derived a LRD mass function from the uSIDM model at $\zobs\sim5$, with a fixed duty cycle $\varepsilon_{\rm{DC}} = 0.01$ and a log-normal distribution of accretion rates based on Eddington ratios with mean $\lambda_{0}=0.2$ and standard deviation $\sigma_{\lambda} = 0.3$, which we reproduce here in Fig.~\ref{fig:uSIDM_LRD_mass_func}. In particular, we derived a ``full mass" function and a ``power-law" mass function. The full mass function includes a toy BH merger model, an AGN duty cycle, and a luminosity cutoff matching observational constraints. The power-law mass function is a simple power-law fit of the form $\philrd(\mBH) = A\mBH^{\alpha}$, where $\log_{10}A$ and $\alpha$ are our normalization and slope that we fit to our full model; this gives an analytic approximation to our full mass function. In Fig.~\ref{fig:uSIDM_LRD_mass_func}, the colored points are the full mass function, whereas the solid and black dashed lines are the median and 1-sigma errors on the power-law fit.

While the duty cycle enters the number density normalization, the effective bias is independent of the duty cycle and is determined solely by the weighted halo mass distribution. Thus, the assumed value of $\varepsilon_{\rm{DC}}$ does not impact our clustering prediction. A duty cycle that depends on halo mass or other secondary halo properties could in principle modify this statement. However, such a dependence is currently unconstrained, so for simplicity we assume that $\varepsilon_{\rm{DC}}$ is constant. Similarly, for our uSIDM model, we do not require accretion at the Eddington limit to grow our BH seeds because uSIDM halos can collapse and form much heavier seeds; hence, we model the Eddington ratios with a mean of $\lambda_{0}=0.2$, which is more typical of high-redshift AGN \citep{fangzhou_LRD,Xiao_2021,Willott_2010}. 

\section{LRD Clustering and the Effective Bias}\label{sec:beff}

The halo bias, $b_h$, is the relation between the two-point dark matter halo correlation function, and the two-point matter correlation function: $\xi_{hh} \simeq b^{2}_{h}\xi_{mm}$. But since we are interested in a specific sub-population of dark matter halos as a tracer for LRDs, we need to calculate an effective bias. Therefore, 
we define the effective halo bias \citep{Porciani:2004vi,Fanidakis:2013vva} as,

\begin{equation}
      \beff =\frac{\int d\log_{10}m_{200}~N_{\rm{LRD}}~n(m_{200})~b_h[m_{200},z]}{\int d\log_{10}m_{200}~N_{\rm{LRD}}~n(m_{200})},
\label{eq:beff_general}
\end{equation}

\noindent where $N_{\rm{LRD}}$ is the mean number of LRDs that are hosted in halos of mass $m_{200}$, $n(m_{200})$ is the number density of DM halos with mass $m_{200}$, and $b_h[m_{200}(\mBH),z]$ is the standard halo bias for which we use \cite{Tinker2010} via \texttt{COLOSSUS}~\citep{colossus}. Equation~\eqref{eq:beff_general} is given in terms of the halo mass, but under a change of variables, one can derive the effective bias in terms of the LRD mass function as a function of $\mBH$, instead of the halo mass function:

\begin{equation}
      \beff =\frac{\int d\log_{10}m_{200}~\philrd(\mBH)~\frac{d\log_{10}\mBH}{d\log_{10}m_{200}}~b_h[m_{200},z]}{\int d\log_{10}m_{200}~\philrd(\mBH)~\frac{d\log_{10}\mBH}{d\log_{10}m_{200}}}.
\label{eq:beff}
\end{equation}

\noindent where $\philrd(\mBH) = dn_{\rm LRD}/d\log_{10}\mBH$ is our LRD mass function, either the power-law fit or the full model.  Physically, the effective bias is the number density weighted average quantity of the large scale LRD clustering strength. Since the halo bias is a monotonically increasing function with $m_{200}$, $\beff$ is a direct probe of the underlying halo environment of LRDs. 

In principle, the LRD occupation may also depend on secondary halo properties associated with assembly bias, such as the concentration or formation history. In our uSIDM formalism, the most relevant assembly bias dependence is the dependence on the concentration, since the collapse timescale depends on both $m_{200}$ and $c_{200}$. Therefore, scatter in the concentration-mass relation will propagate into the collapse time, seed black hole mass, the final black hole mass, and LRD occupation. We separately consider two effects due to halo concentration: (1) the impact of variations in $c_{200}$ on the predicted LRD mass function, $\philrd$, and the resulting $\beff$; and (2) concentration-induced assembly bias in the halo bias itself. We test the former by modifying the concentration-mass relation and the latter using concentration-dependent assembly bias corrections from \cite{Wang_LRD_bias}.

\begin{figure}
	\includegraphics[width=1.05\columnwidth]{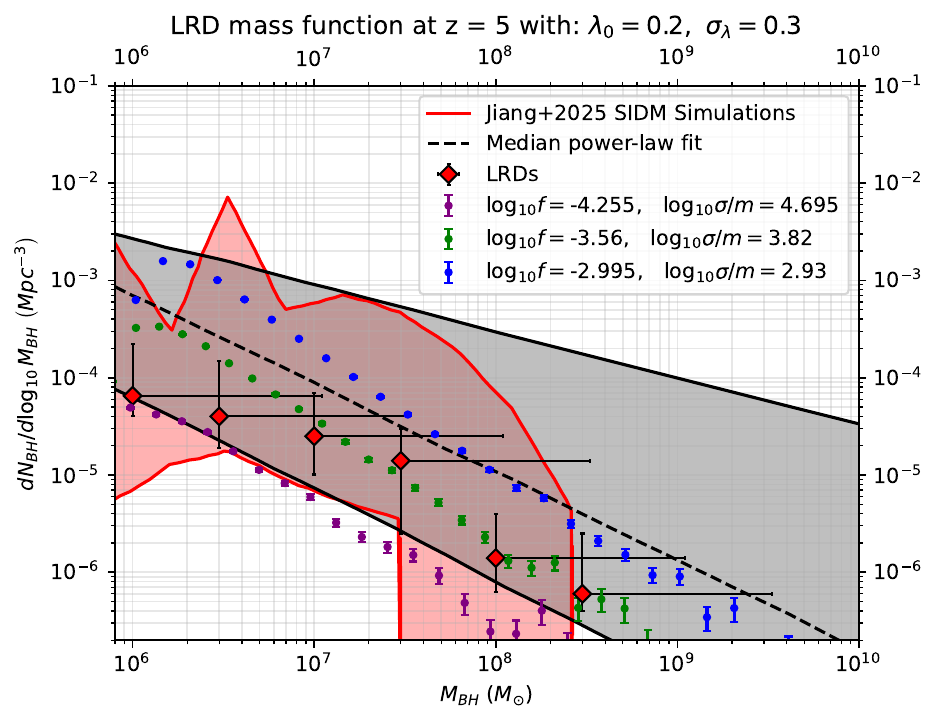}
    \caption{Here we reproduce Fig. 2 (right) from \protect\cite{Roberts_LRD}. The uSIDM LRD mass function (colored dots) for different uSIDM parameters is shown. The uSIDM results are plotted against the SIDM simulation results from \protect\cite{fangzhou_LRD} (shaded red), and the LRD mass function data points from \protect\cite{Kokorev_2024} (red diamonds). The red diamonds are the derived BH masses, assuming $\lambda_{\rm{Edd}} = 1$; for lower accretion rates these would shift to larger BH masses as depicted by the horizontal lines which extend down to $\lambda_{\rm{Edd}} = 0.1$. The simulations for the red shaded region assume $\varepsilon_{\rm{DC}} = 1$ whereas we assume $\varepsilon_{\rm{DC}} = 0.01$.}
    \label{fig:uSIDM_LRD_mass_func}
\end{figure}

\begin{figure}
	\includegraphics[width=1.07\columnwidth]{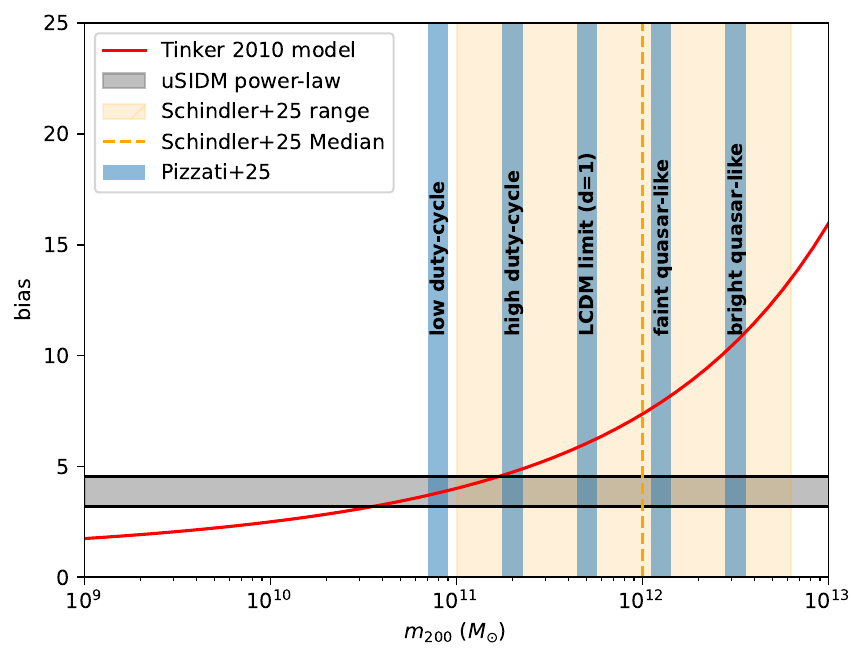}
    \caption{Here we show the range of $\beff$ from the three power-law fits to the uSIDM LRD mass function, $3.20 \lesssim \beff \lesssim 4.55$ (gray shaded region), overlaid onto the \protect\cite{Tinker2010} halo bias function (red line). We find a rough scale for $\mchar$ to be $\lrp{0.36-1.7} \times 10^{11}~\msun$. We also show the inferred mass range from \protect\cite{Schindler+25} in shaded orange, with the median inferred value as the dashed orange vertical line. In blue we plot the various cases explored by \protect\cite{Pizzati+25}.}
    \label{fig:power-law-bias}
\end{figure}

We evaluate the mass integrals over the range of our BH masses from Fig.~\ref{fig:uSIDM_LRD_mass_func}. To get a first approximation to $\beff$, we use a power-law fit to $\philrd(\mBH) = A\mBH^{\alpha}$, where we pick three fiducial values of $\log_{10}A$ and $\alpha$ that represent the solid black and dashed black gray lines in Fig.~\ref{fig:uSIDM_LRD_mass_func}; specifically we use $\{\log_{10}A, \alpha\} = \{2.22,-0.9\}$ for the central fit and $\{\log_{10}A, \alpha\} = \{1.34,-0.92\}$ and $\{0.27,-0.47\}$ for the 1-sigma range. These three fiducial parameter choices yield $\beff=\lrp{3.24,~3.20,~4.55}$.  These results are shown in Fig.~\ref{fig:power-law-bias}, and we overlay the halo bias $b_h[m_{200}(\mBH),z]$ from \cite{Tinker2010} in red. In order to determine the characteristic mass scale, $\mchar$, for these values of $\beff$, we numerically solve $b_h[\mchar(\mBH),z] = \beff$, and we find $\mchar = \lrp{3.79\times10^{10}~\msun,~3.59\times10^{10}~\msun,~1.69\times10^{11}~\msun}$ for the three values of $\beff$ respectively. These three power-laws span a large degree of the uSIDM parameter space, and thus the sensitivity to the slope of the LRD mass function and the implied characteristic mass scale. Therefore, even with a simple power-law fit, the implied $\mchar$ of LRDs fall on the low-to-moderate halo mass range compared to high-redshift luminous quasars as can be seen by the orange and blue bands which are the inferred mass ranges by the single LRD within an overdensity of eight galaxies in \cite{Schindler+25} and the mock-clustering simulations of \texttt{JWST} fields by \cite{Pizzati+25} respectively.

Equation~\eqref{eq:beff} for $b_{\rm{eff}}$ works for a direct one to one mapping between $\mBH$ and $m_{200}$, i.e., for a fixed $\lambda_{\rm{edd}}$. But in general we sample from a log-normal distribution in $\lambda_{\rm{edd}}$, thus we need to sample the integral. To do this, we employ a Monte Carlo procedure which samples from the distributions of $\mBH$ and $m_{200}$ for our full LRD mass function which includes the modeled distribution of accretion and merger histories. Specifically, we take $N = N_{mc}$ samples with Monte Carlo weights $w_{i}$ which randomly select from the final sample of BH masses and halo masses that survive the luminosity cutoff (see \cite{Roberts_LRD} Section 3.2):

\begin{equation}
    b_{\rm{eff}} = \frac{\sum_{i}^{N_{mc}} w_{i}b_h[m_{200,i},z]}{\sum_{i}^{N_{mc}} w_{i{}}} = \frac{1}{N_{mc}}\sum_{i}^{N_{mc}}b_h[m_{200,i},z]\,,
\label{eq:MC-beff}
\end{equation}

\noindent where $w_{i} = w_{0}\varepsilon_{\rm DC}$ and
$w_{0} = \frac{n_{\rm seed}}{N_{mc}}$, such that $n_{\rm seed}$ is the original number density of seeds:

\begin{equation*}
    n_{\rm seed} = \int_{\log_{10}m_{200}^{\rm min}}^{\log_{10}m_{200}^{\rm max}} \frac{dN_{\rm{Halo}}}{d\log_{10}m_{200}}d\log_{10}m_{200}~,
\end{equation*}

\noindent where $\frac{dN_{\rm{Halo}}}{d\log_{10}m_{200}}$ is the halo mass function, for which we use \cite{Tinker2008}. Our mass range for the halos is $10^{7.5}$--$10^{11.5}$ $M_{\odot}$ as these are in the range of most efficient core-collapse within uSIDM \citep{Roberts_uSIDM}. We note that the weights, $w_i$, in Equation~\eqref{eq:MC-beff} explicitly cancel; however, in practice, our sampling from the halo mass function naturally accounts for the differing abundance of host halos as a function of $m_{200}$. Combined with the stochastic Eddington ratios, the merger model, and luminosity selection, this produces the final Monte Carlo realization over which $\beff$ is averaged. As a result, our full mass function clustering prediction incorporates a more realistic mapping between LRDs and their host halos.

\begin{figure}
	\includegraphics[width=1.08\columnwidth]{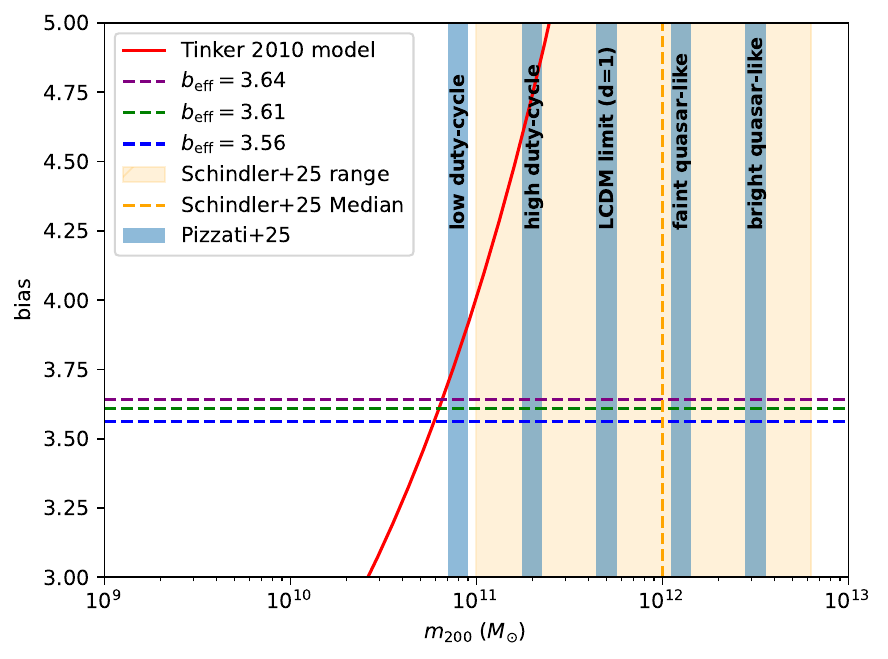}
    \caption{Here we show $\beff$ from the three representative full uSIDM LRD mass function curves. We find that the full mass function yields $\beff$ values that are very nearly insensitive to the underlying particle physics. They all yield $\beff\sim3.6$ and thus $\mchar\sim6.3\times10^{10}~\msun$. 
    To show the small deviations in $\beff$, we zoom into bias values between $3$ and $5$.
    As in Fig.~\ref{fig:power-law-bias}, we plot the inferred mass regions from \protect\cite{Schindler+25} in orange and \protect\cite{Pizzati+25} in blue. We find that we are consistent with the low mass and low duty cycle case in \protect\cite{Pizzati+25}.}
    \label{fig:full-model-bias}
\end{figure}

In Fig.~\ref{fig:full-model-bias}, we plot $\beff$ from our full mass function for each of the three examples in Fig.~\ref{fig:uSIDM_LRD_mass_func}, for which we find $\beff=\lrp{3.64,~3.61,~3.56}$. These three models were found in \citep{Roberts_LRD} to bracket the observed LRD mass function. From these values of $\beff$, we find $\mchar \sim 6.3\times10^{10}~\msun\approx 10^{10.8}~\msun$ across our model curves. Given that $\beff$ is consistent to within 1\%, this implies that the predicted clustering bias and the corresponding characteristic mass of LRDs are relatively insensitive to the underlying particle physics, making this a robust prediction of the uSIDM scenario. To test whether scatter in the concentration-mass relation affects this prediction, we consider several explicit modifications to the calculation. First, we apply an extreme coherent shift in halo concentrations, in which we assign each halo $c_{200} = 3\times c_{200}^{\rm{median}}$ at $\zvir\sim13.5.$ Second, we introduce log-normal scatter around the median concentration-mass relation of $0.1$ dex, $0.33$ dex, and $0.55$ dex. In all cases, the effective bias remains unchanged, with $\beff\sim3.61 \pm 0.03$. The realistic scatter case, $0.1$ dex, has no visible impact on $\philrd$, whereas only the extreme cases of $3\times c_{200}^{\rm{median}}$ and a scatter of $0.55$ dex have any noticeable impact on the normalization of $\philrd$. We therefore conclude that realistic scatter in the concentration-mass relation has a negligible impact on the predicted LRD population and effective bias for the rapid gravothermal evolution and core-collapse uSIDM regime considered here. In Appendix~\ref{sec:appendix-c200-dist}, we also verify that the concentration distribution of our LRD hosting halos closely follow that of the overall halo population, indicating that our model does not preferentially select extreme concentration outliers.

To assess whether concentration-induced assembly bias affects this prediction, we carried out a sensitivity test using the highest and lowest 25\% concentration-quartile assembly bias corrections from \citet{Wang_LRD_bias}. We considered two limiting cases: first, we assumed that all of our LRD host halos occupy the highest concentration quartile and applied the corresponding correction to $\beff$; second, we assumed that the halos occupy the lowest concentration quartile. For our chosen uSIDM parameter cases, these extreme tests changed the inferred $\beff$ by no more than 0.5. Since our LRD host halos do not generally populate these extreme concentration quartiles (Appendix~\ref{sec:appendix-c200-dist}), we expect the actual concentration-induced assembly bias to be a substantially smaller effect. Our semi-analytic model does not track individual halo assembly histories, and therefore does not predict a formation age distribution of the LRD host halo population. However, given that the formation age assembly bias corrections from Wang et al. (2026) are smaller than the concentration-induced assembly bias corrections, we expect the effect of formation age assembly bias on $\beff$ to be correspondingly smaller.

For each of the three example uSIDM parameters used in the full mass function calculation, we also compute the host halo mass distributions directly from the modeled LRD population. The median $m_{200}$ values for these three cases are all approximately $6.7 \times 10^{10}~\msun\approx10^{10.8}~\msun$, which is in good agreement with the characteristic mass scale inferred from $\beff$. This value of $\beff$ suggests that LRDs in this scenario represent a distinct population from high-redshift quasars, which have a much larger bias and characteristic mass scale, as seen by the orange and blue regions in the figure. Therefore, our result is in tension with the mass scale derived in \cite{Schindler+25} - but is consistent with the low-mass and low-duty cycle forecast performed by \cite{Pizzati+25} and the faint AGNs analyzed by \cite{Arita+24} and \cite{Lin+25}.

\section{Summary and Conclusions}\label{sec:conclusions}

In this Letter, we use a model for the LRD mass function resulting from early core-collapse in scenarios where a fraction of the dark matter is ultra-strongly self-interacting to calculate the effective clustering bias, $\beff$, and characteristic mass scale, $\mchar$, of LRDs seeded by uSIDM. We find that both an approximate power law fit to the mass function and a full Monte Carlo sampling of the astrophysical parameters (the accretion and merger rates) are in rough agreement with the predicted LRD clustering. We also find that $\beff$ is insensitive to the underlying particle physics, and hence, our model predictions are robust across the relevant parameter space. While it does not directly constrain the underlying particle parameters, $\beff$ is a uniform, consistent prediction from generic uSIDM seeding LRD formation, and is thus an excellent observational test of this model. Based on the uSIDM value of $\beff\sim3.6$ and hence $\mchar\sim6.3\times10^{10}~\msun$, we infer that LRDs in this scenario would represent a distinct population inhabiting lower-mass halos than high-redshift quasars. Thus, our model predictions are consistent with characteristic masses inferred by faint AGN analysis by \cite{Arita+24} and \cite{Lin+25}, as well as the low-mass low-duty cycle scenario explored by \cite{Pizzati+25}. This distinct prediction is falsifiable and therefore constitutes a clear observational test of the uSIDM seeded LRD formation scenario and can be tested with upcoming \texttt{JWST} surveys. This is, to our knowledge, the first quantitative prediction of the clustering of LRDs, from a formation scenario that maps consistently onto the LRD mass function. 

While we have focused on uSIDM seeded LRD formation, it is worth emphasizing that different physical production mechanisms of BH seeds generically predict different values for $\beff$, due to their distinct assumptions about the mapping between black hole mass and host halo mass, including effects from the concentration-mass relation, uncertainties on initial seed masses, and the subsequent growth history, which together determine the $\mBH$ to $m_{200}$ relation. A systematic comparison of $\beff$ predictions across different LRD formation channels is therefore an important and exciting avenue for future theoretical work. 

In light of this, while our model produces an initial prediction, we can generalize the uSIDM model to include other effects to refine the predicted LRD number densities and the uSIDM $\mBH-m_{200}$ relation, such as adding dark Bondi accretion to our seed growth~\citep{dark_bondi}, testing different log-normal distributions for Eddington accretion rates, different duty cycles, different luminosity cut offs, different BH merger scenarios, and N-body simulations of uSIDM. These extensions may allow us to model LRD growth and abundance more self-consistently, and potentially tighten the mapping between BH and halo mass. With these future refinements, we may also be able to predict the clustering redshift evolution as more LRDs are found at different redshifts. Because LRD clustering is a tracer for early BH formation, growth, and feedback, these measurements may help us to discriminate between heavy and light seed scenarios, and thereby give insight to the physical origin of LRDs.

\section*{Acknowledgements}

This work is partly supported by the U.S.\ Department of Energy grant number de-sc0010107 (SP).

\section*{Data Availability}

No new data were generated or analyzed in support of this research.



\bibliographystyle{mnras}
\bibliography{references} 




\appendix

\section{LRD Hosting Halos Concentration Distribution}\label{sec:appendix-c200-dist}

Here we include the $c_{200}$ distribution for our LRD hosting halos along with the overall host halo distribution, for one set of the uSIDM parameters. We find that our LRD hosting halos are representative of the overall halo population, and do not preferentially occupy the extreme tails of the concentration distribution.

\begin{figure}
    \includegraphics[width=\columnwidth]{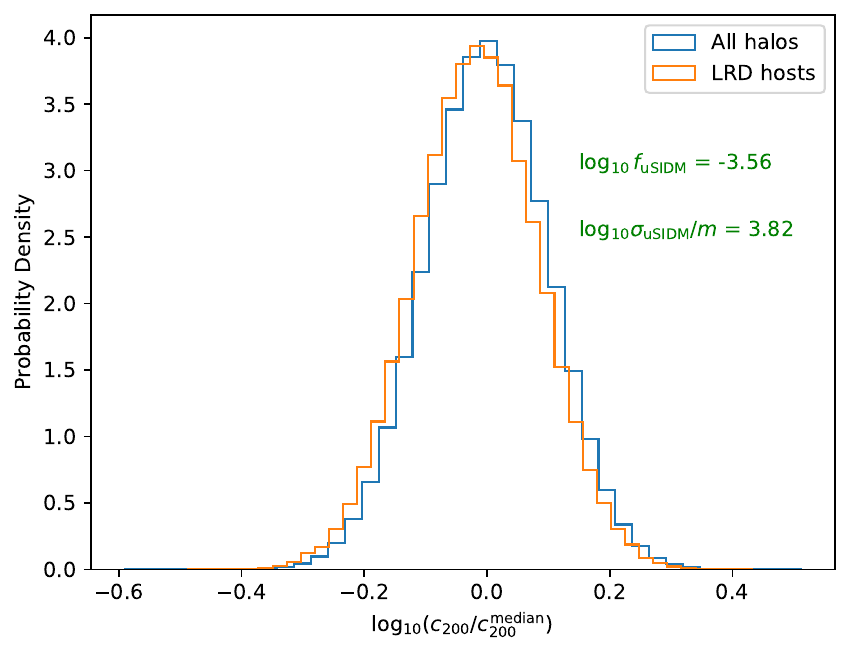}
    \caption{The comparison between the concentration of all of the LRD hosts to the entire halo population from our Monte Carlo, for one of the uSIDM parameter sets.}
\end{figure}


\bsp	
\label{lastpage}
\end{document}